\documentclass[11pt,reqno]{amsart}
\usepackage{graphicx,amsmath,amsfonts,latexsym,amssymb,amsthm,color}
\usepackage{latexsym}
\usepackage{verbatim}
\usepackage{enumerate}
\usepackage{enumitem}

\renewcommand{\Re}{\operatorname{Re}}
\renewcommand{\Im}{\operatorname{Im}}

\newcommand{\der}{\mathrm{d}}
\newcommand{\rmi}{\mathrm{i}}

\newcommand{\tr}{\mathrm{Tr}}

\newcommand{\rel}{\mathrm{rel}}

\newcommand{\R}{\mathbb{R}}
\newcommand{\C}{\mathbb{C}}

\newcommand{\Z}{\mathbb{Z}}

\setlist{leftmargin=8mm}

\newtheorem{theorem}{Theorem}[section]
\newtheorem{definition}[theorem]{Definition}

\title[Dimensional reduction formulae]{Dimensional reduction formulae for spectral traces and Casimir energies}

\author[A.~Strohmaier]{Alexander Strohmaier}
\address{Leibniz University Hannover, Institute of Analysis, 30167 Hannover, Germany}  \email{a.strohmaier@math.uni-hannover.de}

\begin{document}

\begin{abstract}
 This short letter considers the case of acoustic scattering by several obstacles in $\R^{d+r}$ for $r,d \geq 1$ of the form $\Omega \times \R^r$, where $\Omega$ is a smooth bounded domain in $\R^d$. As a main result a von-Neumann-trace formula for the relative trace is obtained in this setting. As a special case we obtain a dimensional reduction formula for the Casimir energy for the massive and massless scalar fields in this configuration $\Omega \times \R^r$ per unit volume in $\R^r$.
 \end{abstract}

\maketitle

\section{Introduction: Casimir energy as a relative trace}

In this section I briefly review results on the relative trace, as introduced in \cite{RTF}, and the relation to the Casimir energy.
The setting is as follows. We consider a smooth bounded open subset $\Omega \subset \R^d, d \geq 2$ which consists of connected components $\Omega_1,\ldots,\Omega_N$. For notational simplicity we assume further that $\R^d \setminus \Omega$ is connected. 
We will think of $\Omega$ as a collection of objects $\Omega_j$ that we refer to as obstacles. The boundary $\partial \Omega$ is a codimension one submanifold with connected components $\partial \Omega_1,\ldots,\partial \Omega_N$. The complement $E=\R^d \setminus \overline{\Omega}$ will be referred to as the exterior region. There are a variety of Laplace-operators one can associate to this configuration.

\begin{definition}
We define the unbounded densely defined self-adjoint operators  $\Delta_0, \Delta_j, \Delta$ on the Hilbert space  $L^2(\R^d)$ as follows.
\begin{enumerate}
	 \item the free Laplacian $\Delta_0$ is defined as the self-adjoint operator $\Delta$ with domain $H^2(\R^d) \subset L^2(\R^d)$.
 \item   the operator $\Delta_{j}$ is defined as the self-adjoint operator associated to the Dirichlet quadratic form
on $H^1_0(\R^d \setminus \partial \Omega_j)$, i.e. is the Laplace operator with Dirichlet boundary conditions imposed on $\partial \Omega_j$. 
 \item   the operator $\Delta$ is defined as the self-adjoint operator associated to the Dirichlet quadratic form
on $H^1_0(\R^d \setminus \partial \Omega)$, i.e. is the Laplace operator with Dirichlet boundary conditions imposed on $\partial \Omega$. This operator is the direct sum of interior and exterior Laplace operators $\Delta = \Delta_\mathrm{int} \oplus \Delta_\mathrm{ext}$ on $L^2(\R^d) = L^2(\Omega) \oplus L^2(E)$.
\end{enumerate}
\end{definition}

We fix $m \geq 0$.
The main result of \cite{RTF} implies that for any $s>0$ the operator
$$
 D_s = \left( (\Delta+m^2)^{\frac{s}{2}} - (\Delta_0+m^2)^{\frac{s}{2}} \right)  - \sum_{j=1}^N \left((\Delta_j+m^2)^\frac{s}{2} - (\Delta_0+m^2)^\frac{s}{2} \right)
$$
is trace-class in the sense that it is bounded and its unique continuous extension to all of $L^2(\R^d)$ is trace-class. Its trace is referred to as the relative trace, as it captures the contributions to the trace that are present because of the influence of the individual objects on each other.
This relative trace can be computed by the formula (\cite{RTF}, Theorems 1.6 and 1.7)
$$
 \mathrm{tr}(D_s) = \frac{s}{\pi} \sin(\frac{\pi}{2} s) \int_m^\infty \lambda (\lambda^2 - m^2)^{\frac{s-2}{2}} \Xi(\rmi \lambda) \der \lambda. 
$$
Here the function $\Xi$ is defined as a Fredholm determinant
$$
 \Xi(\lambda) = \log \det \left( S_\lambda \circ S_{\lambda,\mathrm{diag}}^{-1} \right),
$$
where $S_\lambda: L^2(\partial \Omega) \to L^2(\partial \Omega)$ is the Helmholtz single layer operator, and 
$S_{\lambda,\mathrm{diag}} = \oplus_{j=1}^N S_{\lambda,j}$ is the direct sum of the Helmholtz single layer operators of the individual objects. The Helmholtz single layer operator associated to the domain $\Omega$ is a standard boundary layer operator (see for example the monograph \cite{McLean2000}). For continuous $f\in C(\partial \Omega)$ it is explicitly given in terms of the weakly singular integral
\begin{align*}
 (S_\lambda f)(x) = \int_{\partial \Omega} G_\lambda(x,y) f(y) \der y,
\end{align*}
where $G_\lambda$ is the Green's function of the Helmholtz equation, i.e. the distributional integral kernel of
$(\Delta_0 - \lambda^2)^{-1}$. For fixed $\lambda \not=0$ the operator $S_\lambda$ is a pseudo-differential operator of order $-1$. It is an analytic function in the variable $\lambda$ on the upper half plane that extends analytically to a logarithmic cover of the complex plane.

The function $\Xi(\lambda)$ is
 analytic in a neighborhood of the positive imaginary axis and exponentially decaying as the imaginary part goes to infinity. For $s=1$ the quantity $\frac{1}{2}\mathrm{tr}(D_s)$ has the interpretation of the Casimir energy of the configuration of objects for the scalar field of mass $m$ and Dirichlet boundary conditions. This interpretation was rigorously justified in \cite{MR4407742}, where it was shown that its directional derivative coincides with the force as computed from the normally ordered stress energy tensor. The definition of the Casimir energy as a relative trace does not require regularisation and identifies it as the trace of a trace-class operator.

This leads to the formula
$$
 E_\mathrm{Cas} = \frac{1}{2 \pi} \int_m^\infty \frac{\lambda}{\sqrt{\lambda^2 - m^2}} \Xi(\rmi \lambda) \der \lambda,
$$
which is very useful for numerical computations and also for obtaining asymptotics. We note here that this is also valid in dimension $d=1$, with some obvious modifications as the boundaries just becomes points.

Formulae as such for the Casimir energy have at least for $m=0$ been obtained and derived in the physics literature by means of path integrals describing surface fluctuations (\cite{kenneth06, EGJK2007, emig2008casimir, kenneth08, milton08, johnson2011numerical}), which can be understood as the mechanism behind the Casimir effect.
On the mathematical side related determinant formulae have also been obtained in the context of scattering theory and the spectral shift function \cite{MR0139007} which in many way generalises the eigenvalue counting function (\cite{MR512084, Majda_1978, MR956828}).
The spectral shift function is not the same as the function $\Xi$, but is related to via a linear tranformation (see \cite{MR4396069}) that allows to relate it to the Duistermaat-Guillemin-Gutzwiller trace formula. In the context of the spectral shift function a determinant formula  has been obtained by Carron in \cite{carron1999determinant}.

In this paper we will drop the assumption of compactness of the obstacles but instead consider objects  that are translation invariant in certain directions. We assume therefore that we have extra $r$-many dimensions in which the objects are translation invariant. In other words, with the same notations as before, we consider the sets $\tilde \Omega_j = \Omega_j \times \R^r$ and their union
$\tilde \Omega = \Omega \times \R^r$. The exterior regions is then $\tilde E = E \times \R^r$, and the corresponding operators
$\tilde \Delta,\tilde \Delta_j, \tilde \Delta_0$ are defined in the same way as before, with $\Omega$ replaced by $\tilde \Omega_j$.
We note that these operators can be written as
\begin{align*}
 \tilde \Delta &= \Delta \otimes 1 + 1 \otimes \Delta_t,\\
  \tilde \Delta_j &= \Delta_j \otimes 1 + 1 \otimes \Delta_t,\\
   \tilde \Delta_0 &= \Delta_0 \otimes 1 + 1 \otimes \Delta_t,\\
\end{align*}
with respect to the decomposition $L^2(\R^{d+r}) = L^2(\R^{d}) \hat \otimes L^2(\R^{r})$, and $\hat \otimes$ denotes as usual the completed tensor-product of Hilbert spaces.. Here $\Delta_t$ is the transversal Laplace, which is the free Laplace operator on $L^2(\R^r)$. 

We can then define the operator
$$
 \tilde D_s = \left(  (\tilde\Delta+m^2)^{\frac{s}{2}} - ( \tilde \Delta_0+m^2)^{\frac{s}{2}} \right)  - \sum_{j=1}^N \left( (\tilde\Delta_j+m^2)^\frac{s}{2} - (\tilde \Delta_0+m^2)^\frac{s}{2} \right).
$$
Since this operator is translation invariant with respect to the $\R^r$ action it is only trace-class if it vanishes. The latter is typically not the case. One can show that $\tilde D_s$ has a smooth integral kernel on $(\tilde E \cup \tilde \Omega)^2$, and the restriction to the diagonal is an $\R^r$-invariant function on $\R^{d+r}$. This can be integrated over $\R^d$ to yield what is often referred to as the von-Neumann trace.

\begin{theorem} \label{mainth}
The operator $\tilde D_s$ is trace-class with respect to the von-Neumann trace $\widetilde{\mathrm{tr}}$ for any $m\geq 0, s>0$. The corresponding trace equals
$$
 \widetilde{\mathrm{tr}}(\tilde D_s) = -\frac{2^{1-r} \pi ^{-r/2}}{\Gamma \left(-\frac{s}{2}\right) \Gamma
   \left(\frac{r+s}{2}\right)} \int_m^{\infty} \lambda (\lambda^2 - m^2)^{\frac{s+r}{2}-1}\Xi(\mathrm{i} \lambda) d \lambda.
$$
\end{theorem}
For $s=1$ this formula can be used to compute the Casimir energy per unit volume in $\R^r$.
I would like to emphasise here that $D_s$ is usually not a non-negative operator. Therefore the statement that the operator is in the trace-ideal with respect to the Neumann trace is stronger than an integrability statement of the diagonal of the integral kernel.

Whereas for notational convenience we restricted ourself to the case $d \geq 2$
the formula, with the obvious modifications, also applies to the case $d=1$. It then yields the classical formula for the Casimir energy. Two parallel Dirichlet plates indeed correspond to two points in $\R$ that are multiplied by $\R^2$, so we obtain infinitely extended parallel plates in $\R^3$. 
In case $d=1$, $r=2$, $m=0$, $s=1$ Theorem \ref{mainth} specialises to
$$
\tilde E_\mathrm{Cas}=  \frac{1}{2}\widetilde{\mathrm{tr}}(\tilde D_1) = \frac{1}{4 \pi^2} \int_0^\infty \lambda^2 \Xi(\mathrm{i} \lambda) \der \lambda.
$$
In one dimension the free Green's function is given by $G_\lambda(x,y) = -\frac{1}{\rmi \lambda} e^{\rmi \lambda |x-y|}$, which yields
$$
 S_\lambda =  -\frac{1}{\rmi \lambda} \left( \begin{matrix} 1 &  e^{\rmi \lambda |x-y|}\\ e^{\rmi \lambda |x-y|} & 1\end{matrix} \right), \quad S_{\lambda,\mathrm{diag}} =  -\frac{1}{\rmi \lambda} \left( \begin{matrix} 1 &  0\\ 0 & 1\end{matrix} \right),
$$
for the matrix of single layer operators.
Therefore, one computes for the corresponding Fredholm determinant $\Xi(\mathrm{i} \lambda) = \log (1 - e^{-2 \lambda a})$, where $a$ is the distance between the two plates. This gives a convergent integral, namely
$$
 \tilde E_\mathrm{Cas} = \frac{1}{4 \pi^2} \int_0^\infty \lambda^2 \log (1 - e^{-2 \lambda a}) \der \lambda = - \frac{\pi^2}{1440 a^3}
$$
the classical formula for the Casimir energy per unit volume for the massless free scalar field between two plates of the form $\{\mathrm{pt}\} \times \R^2$ with Dirichlet boundary conditions.
The formula for the actual electromagnetic Casimir effect differs from this one by a factor of two for the plate configuration. For a general treatment of the electromagnetic field one requires boundary layer methods adapted to Maxwell's equations as explained in \cite{OSM}. Dimensional reduction formulae are also expected for Maxwell's equations, but require a boundary layer analysis of the Laplace operator on one forms with relative boundary conditions. 

Dimensional reduction formulae for the Casimir energy within the context of zeta regularisation of spectral traces are not new and can be found throughout the literature on the Casimir effect. The key novelty in the present approach via relative operators is the inherent finiteness and the aspect that the operators are trace-class in the correct von-Neumann trace-ideal.

\section{Neumann trace and dimensional reduction}

Recall that given a von-Neumann algebra $\mathcal{N}$ a {\sl trace} is a positively homogeneous map $\tau$ from its positive cone $\mathcal{N}_+$ to $\overline{\R_+}=[0,\infty) \cup \{\infty\}$ satisfying $\tau(a^* a) = \tau(a a^*)$. The $p$-th {\sl Schatten ideal} can be defined as the set of elements $a$ with $\tau(|a|^p) < \infty$.

I would briefly like to review the notion of the {\sl von-Neumann trace} as introduced by von-Neumann and used in many contexts in mathematics, for example the Novikov conjecture and Atiyah's $L^2$-index theorem. For this consider an operator on the Hilbert space $L^2(\R^{d+r}) = L^2(\R^{d}) \hat\otimes L^2(\R^{r})$.  If a discrete group $\Gamma$ acts isometrically on $\R^{r}$
this action induces an action on $L^2(\R^{d+r})$. Here we will simply choose $\Gamma = \Z^r$ which acts by translations. In this case the unit-cube $[0,1]^r$ is a fundamental domain for this action.
We now consider the von Neumann algebra $\mathcal{A}_\Gamma$ of bounded operators $A$ that commute with the action of $\Z^r$. Using the fundamental domain we have the decomposition 
$L^2(\R^{d+r}) = L^2(\R^{d}) \hat\otimes L^2([0,1]^{r}) \hat\otimes \ell^2(\Gamma)$. The algebra $\mathcal{A}_\Gamma$ is then isomorphic to the algebra of bounded operator on the Hilbert space $\mathcal{H}=L^2(\R^{d}) \hat\otimes L^2([0,1]^{r})$ via the map
$$
 \mathcal{L}(\mathcal{H}) \to \mathcal{A}_\Gamma, \quad A \mapsto A \otimes 1.
$$
The standard trace for bounded operators on $L^2(\R^{d}) \hat\otimes L^2([0,1]^{r})$ induces a trace $\mathrm{tr}_\Gamma$ on the von-Neumann algebra $\mathcal{A}_\Gamma$.
By Mercer's theorem, if $A \in \mathcal{A}_\Gamma$ is in $\mathcal{A}_\Gamma^1$, has a continuous integral kernel
$a(x,y,x',y')$, then $\Gamma$-trace can be computed as
$$
 \mathrm{tr}_\Gamma(A) =\int_{(0,1)^r} \left(\int_{\R^d} (a(x,x,y,y) \der x \right)\der y.
$$

The algebra of $\Gamma$-invariant operators contains of course the sub-algebra of $\R^m$-invariant operators, i.e. operators which are translation invariant with respect to the $\R^m$-action. We will call this algebra $\mathcal{A}_0$. The trace $\mathrm{tr}_\Gamma$ then restricts to a trace $\mathrm{tr}_0$ on $\mathcal{A}_0$. 

I now give an explicit description of this trace and the corresponding trace-ideal in $\mathcal{A}_0$.
We can use the Fourier transform on $\R^m$ to see that translation invariant operators on $L^2(\R^m)$ are Fourier multipliers.
Using a partial Fourier transform $\tilde{\mathcal{F}}=1 \otimes \mathcal{F}$ on $L^2(\R^{d}) \hat \otimes L^2(\R^{r})$ we see that the mapping $\mathcal{A}_0 \to \mathcal{B}(L^2(\R^{d})) \hat\otimes \mathcal{B}(L^2(\R^{r})), A \mapsto \tilde{\mathcal{F}}^{-1} A \tilde{\mathcal{F}}$ restricts to an isomorphism $\mathcal{A}_0  \to \mathcal{B}(L^2(\R^{d})) \hat\otimes L^\infty(\R^{d}) =  L^\infty(\R^{r},  \mathcal{B}(L^2(\R^{d}))$. We will use this isomorphism to identify $\mathcal{A}_0$ with $L^\infty(\R^{d},  \mathcal{B}(L^2(\R^{d}))$. The positive cone is given by elements $A(\xi)$ such that $A \geq 0$ almost everywhere.
The $p$-th Schatten ideal is the set
$$
 \mathcal{A}^p_0 = \{A \in L^\infty(\R^{d},  \mathcal{B}(L^2(\R^{d})) \mid \int_\R \mathrm{tr}(|A(\xi)|^p) \der \xi < \infty \}.
$$
The trace on  $\mathcal{A}^1_0 $ then given by
$$
 \tr_0(A) = \frac{1}{(2 \pi)^r} \int_\R \mathrm{tr}(A(\xi)) \der \xi.
$$

Indeed, we can consider the corresponding Laplace operator $\tilde \Delta$ on $\R^{m+r}$ with Dirichlet boundary conditions on $\R^r \times \partial \Omega$. Similarly, we can consider the operators $\tilde \Delta_j$ and $\tilde \Delta_0$. Then the integral kernel of the operator
$$
 \tilde D_f = f(\tilde \Delta^{\frac{1}{2}}) - f(\tilde \Delta_{0}^{\frac{1}{2}}) - \left(\sum_{j = 1}^{N}[f(\tilde \Delta_{j}^{\frac{1}{2}}) - f(\tilde \Delta_{0}^{\frac{1}{2}})]\right)
$$  
is smooth on $(\R^{m+r} \setminus \partial \Omega) \times (\R^{m+r} \setminus \partial \Omega)$ and the restriction to the diagonal is independent of the $r$-variables. The restriction to the diagonal can then be integrated over $\R^d$. The result of this integration is denoted by $\widetilde{\mathrm{tr}}(\tilde D_f)$. This can be understood as the trace per unit volume in the $\R^r$-direction. We note that $\tilde \Delta$ is a tensor product $\tilde \Delta = \Delta \otimes 1 + 1 \otimes \Delta_{\R^r}$, and a similar formula holds for $\tilde \Delta_j$ and $\tilde \Delta_0$. 
 
\section{Trace-norm estimate}

Following \cite{RTF} we define the relative resolvent
$$
R_\mathrm{rel}(z) = (\Delta  - z^2)^{-1} - (\Delta_0 - z^2) - \sum_{j=1}^N \left((\Delta_j  - z^2)^{-1} - (\Delta_0  - z^2)\right).
$$
By Theorem 4.1 in  \cite{RTF} the relative resolvent is a trace-class operator and the trace-norm $\| R_\mathrm{rel}(z) \|_1$ can be estimated in any sector of the complex plane of the form $S_c =\{z \in \C \mid \Im{z} > c |z|\}, \; c>0$ by
$$
  \| R_\mathrm{rel}(z)  \|_1 \leq C \rho(|z|) e^{- \delta' |z|}.
$$ 

\begin{align}\label{rho}
  \rho(t) :=
 \left\{ 
 \begin{array}{llc}
   t^{d-4}       & \text{ if } d = 2,3 & \\
   |\log t | + 1 & \text{ if } d = 4   & \text{ and } 0 \leq t \leq 1\\
   t^0 = 1       & \text{ if } d \geq 5 & \\
   1             &                      &  \text{ for } t > 1.
 \end{array}  \right.
\end{align}

\begin{figure}[h] 
	\centering
	\includegraphics[scale=1.7]{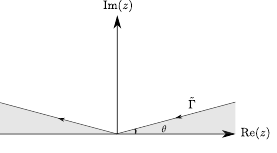}
	\caption{The contour $\tilde{\Gamma}$. The shaded region is mapped to a sector containing the spectrum of $\Delta$ under the map $z \mapsto z^2$.}
	\label{contfig}
\end{figure}

Consider the family of operators $d_s(\xi)$, $\xi \in \R^r$ defined by
\begin{align*}
 d_s(\xi) &= (\Delta + m^2 + \xi^2)^\frac{s}{2} -  (\Delta_0 + m^2 + \xi^2)^\frac{s}{2} \\ &- \sum_{j=1}^N \left( (\Delta_j + m^2 + \xi^2)^\frac{s}{2} -  (\Delta_0 + m^2 + \xi^2)^\frac{s}{2}   \right).
\end{align*}
We can think of this as a family in $L^\infty(\R^r, \mathcal{B}(L^2(\R^d)))$ and the partial Fourier transform shows that this operator corresponds to $\tilde D_s$ under the isomorphism described above.
We have the following representation (see \cite{RTF}, Section 4)
$$
 d_s(\xi) = \frac{\rmi}{\pi} \int_{\tilde \Gamma} \lambda \left( (\lambda^2 + m^2 + \xi^2)^\frac{s}{2} - (m^2 + \xi^2)^\frac{s}{2}\right) R_\rel(\lambda) \der \lambda,
$$
where the integral converges in the Banach space of trace-class operators on $L^2(\R^d)$. This formula is a consequence of the usual formulae one has for the functional calculus of certain sectorial operators (see for example \cite{Haase}) by integrating along the boundary of a sector containaing the spectrum of the operator.
In our case the integration contour $\tilde \Gamma$ is chosen as follows for an arbitrary angle $\theta \in (0,\pi/2)$.
The first part of the contour is the ray $z = -r e^{\rmi \theta}, r \in (-\infty,0]$, and the second part is the ray
$z = -r e^{\rmi (\pi-\theta)}, r \in [0,\infty)$ (see Figure \ref{contfig}).

If $m>0$ the integral converges for all $s \in \C$ and defines a holomorphic family of trace-class operators.
The function $(\lambda^2 + m^2 + \xi^2)^\frac{s}{2}$ has a branch cut at $\lambda = \rmi \sqrt{m^2 + \xi^2}$ and the contour can therefore be moved 
to give 
$$
 d_s(\xi) = \frac{\rmi}{\pi} \int_{\tilde \Gamma+\rmi \delta} \lambda \left( (\lambda^2 + m^2 + \xi^2)^\frac{s}{2} - (m^2 + \xi^2)^\frac{s}{2}\right) R_\rel(\lambda) \der \lambda,
$$
for any $0<\delta<\sqrt{m^2 + \xi^2}$. Using the bound 
on $\| R_\mathrm{rel}(\lambda)\|_1$ we obtain that
$d_s(\xi)$ is trace class for each $\xi$ and 
$$
 \| d_s(\xi) \|_1 \leq C_s e^{-\delta' \sqrt{m^2 + \xi^2}}.
$$
for some $\delta'>0$. Since this is integrable in $\xi$ we have shown that $\tilde D_s$ is in the von-Neumann trace-ideal.
If $m>0$ this also shows that $D_s$ is a holomorphic family of trace-class operators near the real line. For $m=0$ we have an analytic family of trace-class operators in a neighborhood of the positive real line. Furthermore, in case $s>0$ the family is continuous as a function of $m$ at $m=0$.

\section{Proof of the main theorem}

We are now in a position to prove the equality in the main theorem. We first prove the case $m>0$. In this case $\widetilde{\mathrm{tr}}(\tilde D_s)$ is holomorphic near the real line.
We have
$$
 \widetilde{\mathrm{tr}}(\tilde D_s) = \frac{1}{(2 \pi)^r} \int_{\R^d} \tr (d_s(\xi) ) \der \xi.
$$
On the other hand we have
$$
 \tr (d_s(\xi) ) =\frac{\rmi}{\pi} \int_{\tilde \Gamma} \lambda \left((\lambda^2 + m^2 + \xi^2)^{\frac{s}{2}} - (m^2 + \xi^2)^\frac{s}{2} \right) \mathrm{tr}(R_\mathrm{rel}(\lambda)) \der \lambda. 
$$
For sufficiently negative $\Re s$ we have
$$
 \int_{\R^r}  (\lambda^2+ m^2 + \xi^2)^\frac{s}{2} d \xi = \pi^{\frac{r}{2}} \frac{\Gamma(-\frac{s}{2}-\frac{r}{2})}{\Gamma(-\frac{s}{2})} (\lambda^2 + m^2)^{\frac{r+s}{2}}.
 $$
 Thus, for $\Re s$ sufficiently small we can integrate this expression in $\xi$, using Fubini's theorem, we obtain
 $$
  \widetilde{\mathrm{tr}}(\tilde D_s) =  \frac{1}{(2 \pi)^r} \pi^{\frac{r}{2}} \frac{\Gamma(-\frac{s}{2}-\frac{r}{2})}{\Gamma(-\frac{s}{2})} \frac{\rmi}{\pi}\int_{\tilde \Gamma} \lambda  \left( (\lambda^2 + m^2)^{\frac{r+s}{2}} -m^{r+s}\right ) R_\rel(\lambda) \der \lambda.
 $$
Since both sides of the equation are meromorphic they must be equal for all $s \in \C$ when $m>0$. Now assume $s>0$. 
Then the contour can be deformed to be around the imaginary axis and one then obtains 
 the representation
$$
\widetilde{\mathrm{tr}}(\tilde D_s)=   \frac{\pi^{\frac{r}{2}}}{(2 \pi)^r}  \frac{\Gamma(-\frac{s}{2}-\frac{r}{2})}{\Gamma(-\frac{s}{2})} \frac{2}{\pi} \sin(\frac{\pi}{2} (r+s)) \int\limits_m^\infty \lambda \left(\lambda^2-m^2\right)^\frac{r+s}{2} \tr(R_\mathrm{rel}(\rmi \lambda))  \der \lambda.
$$
Finally, using the relation $\tr(R_\mathrm{rel}(\lambda) )= -\frac{1}{2 \lambda} \Xi'(\lambda)$ (\cite{RTF}, Equ.~(30)) and integration by parts gives
\begin{align*}
\widetilde{\mathrm{tr}}(\tilde D_s)=  \frac{\pi^{\frac{r}{2}}}{(2 \pi)^r}  &\frac{\Gamma(-\frac{s}{2}-\frac{r}{2})}{\Gamma(-\frac{s}{2})} \frac{r+s}{\pi} \sin(\frac{\pi}{2} (r+s)) \int\limits_{m}^\infty \lambda \left(\lambda^2-m^2\right)^{\frac{r+s}{2}-1} \Xi(\rmi \lambda)  \der \lambda
\end{align*}
in case $s>0$.
Euler's reflection formula for the $\Gamma$-function then results in
\begin{align*}
\widetilde{\mathrm{tr}}(\tilde D_s)=  -\frac{2^{1-r} \pi ^{-r/2}}{\Gamma \left(-\frac{s}{2}\right) \Gamma
   \left(\frac{r+s}{2}\right)} \int\limits_{m}^\infty \lambda \left(\lambda^2-m^2\right)^{\frac{r+s}{2}-1} \Xi(\lambda)  \der \lambda
\end{align*}
as claimed in the theorem.
The case $m=0$ following by taking limits.

\section*{Conflict of interest and data availablility statement}

The author states that there is no conflict of interest. There is no associated data with this manuscript.

\end{document}